\documentclass[preprint,pre]{revtex4}
\usepackage[T1]{fontenc}
\usepackage[latin1]{inputenc}
\usepackage[dvips]{graphicx}
\usepackage{float}
\usepackage{amssymb}
\usepackage{amsmath}

\newcommand{\lav}{\left\langle}
\newcommand{\rav}{\right\rangle}

\restylefloat{figure}

\begin{document}
\title{Boundary-induced bulk phase transition and violation of Fick's law
in two-component single-file diffusion with open boundaries}
\author{Andreas Brzank\textsuperscript{1}, Gunter M. Schütz\textsuperscript{2}}
\affiliation{\textsuperscript{1}Fakultät für Physik und Geowissenschaften, Universität Leipzig, Abteilung Grenzflächenphysik, Linnestrasse 5, D-04103 Leipzig, Germany \\
\textsuperscript{2}Institut f\"ur Festk\"orperforschung, Forschungszentrum J\"ulich,
52425 J\"ulich, Germany
}

\email{brzank@uni-leipzig.de}
\date{\today}

\begin{abstract}
We study two-component single-file diffusion inside a narrow channel that at 
its ends is open and connected with particle reservoirs. Using a two-species 
version of the symmetric simple exclusion process as a model, we propose a 
hydrodynamic description of the coarse-grained dynamics with a self-diffusion 
coefficient that is inversely proportional to the length of the channel. The 
theory predicts an unexpected nonequilibrium phase transition for the bulk 
particle density as the external total density gradient between the reservoirs
is varied. The individual particle currents do not in general satisfy Fick's 
first law. These results are confirmed by extensive dynamical Monte-Carlo 
simulations for equal diffusivities of the two components.
\end{abstract}

\maketitle

\section{Introduction}

One-dimensional exclusion processes belong to the most studied models in
non-equilibrium statistical mechanics \cite{Ligget85,Schu2001}. Their applications
are manifold. Among others, the symmetric exclusion process (SEP) plays a role in
diffusion where particles, confined in a narrow tube, are not 
allowed to pass each other \cite{Karg92}. This kind of diffusive restriction is 
referred to as single-file diffusion and differs qualitatively 
from normal diffusion described by Fick's law. Whereas in the latter case the mean-square 
displacement of a single particle grows proportional to time, diffusion is 
much slower in the single-file case due to mutual blocking of the particles. The mean-square
displacement grows (for late times) proportional to the square root of time. 
The anomalous behaviour of the mean-square displacement usually serves as 
an experimental indication for the occurrence of single-file diffusion. This requires to trace a single or
more particles which implies to label a certain subset of particles without changing the 
diffusion properties. This corresponds to having a two-species particle system 
with identical diffusion coefficients \cite{Kaerg2002}. Single-file
diffusion is a generic phenomenon observed many years ago for 
molecules diffusing in the channels of certain zeolites 
\cite{Kukla1996}. More recently, single-file behaviour has been demonstrated
in the transport of colloidal spheres confined in one-dimensional channels 
\cite{Wei2000}. Moreover confined 1D random motion plays a role 
in narrow carbon nano tubes, in biological systems like molecular motors or in non-physical systems
such as automobile traffic flow \cite{Schu2003}. Also the famous repton model by Rubinstein and 
Duke \cite{Rubi87,Duke89,Bark96,Leeuwen2003} for the motion of single polymer 
chains, is a lattice gas model of 
this kind. Further motivation for employing single-file diffusion with multiple species comes 
from recent two-species measurements in zeolites \cite{Snurr2002}. Here, a mixture of toluene and
propane was adsorbed into different zeolites. The authors measured the
temperature dependent outflow and noticed a trapping effect, i.e. in a couple
of zeolites the stronger adsorbed toluene molecules influence and control
the outflow of propane.
 
In \cite{Keil2000} the authors review the Maxwell-Stefan theory describing
the diffusive behaviour of a binary fluid mixture where the total current, i.e.
the sum of both species, is zero. The particle-particle
interaction is taken into account by including a friction between 
the species being proportional to the differences in the velocities. This 
approach does not apply to single-file diffusion in a finite system where,
as shown below in the framework of the symmetric simple exclusion process (SEP), 
the self-diffusivity of single particles plays an important 
role in the description of the macroscopic behaviour.

The SEP with one species of particles where classical particles with hard-core
repulsion diffuse on a finite lattice is well understood 
\cite{Spit70,Spoh83,vanB83,Schu94,Ligg99,Schu2001}. At both ends the chain
is connected to a particle reservoir. One is interested in 
stationary-state properties like the density profile determined by the 
reservoir or the stationary particle current as well as the time evolution 
of the particle density and relaxation towards the stationary state.
Our approach for an adequate description of the two-species SEP is a master 
equation description from which we derive an ansatz of coupled
partial differential equations for the macroscopic density profile.

\begin{figure}
\centerline{\includegraphics[width=14cm]{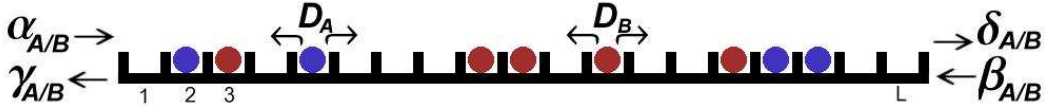}}
\caption{Three-state symmetric exclusion model with open boundaries.}
\label{latticeThreeState}
\end{figure}

We consider a one-dimensional lattice with $L$ lattice sites (Fig.~\ref{latticeThreeState}).
Each site $i$ can be empty or occupied by a particle of type A or B.
Due to hard-core interaction any site carries at most one particle. Particles
can hop to nearest neighbour sites (provided the target site is
empty) according to the constant hopping rates $D_{A/B}$. Hence, $D_A$ ($D_B$) is the
probability of an A (B) particle to attempt a jump per unit time. The model is
defined by random sequential update which forbids simultaneous hopping events.
Let $a_i$ ($b_i$) count the A (B) particles on site $i$. Then the
densities are the expectation values of the respective counters:
$\lav a_i \rav \equiv \rho_A(i)$, $\lav b_i \rav \equiv \rho_B(i)$. The probability of
finding no particle at site $i$ is $\lav v_i \rav=\lav 1-a_i-b_i \rav$.
When we consider a chain with open boundary conditions, particles are
injected and removed according to the boundary rates $\alpha_{A/B}, \gamma_{A/B},
\beta_{A/B}$ and $\delta_{A/B}$, following the notation of \cite{Schu2001}.
The attempt probability per unit time for an A particle to enter the system 
at the left boundary is $\alpha_A$. It leaves the channel at the left boundary according
to $\gamma_A$ as illustrated in Fig.~\ref{latticeThreeState}. The other boundary rates are
defined similarly.

By writing this process in terms of a quantum Hamiltonian formalism
\cite{Schu2001}, the system evolves
in time according to the master equation 
\begin
{align}
 \frac{d}{dt}|P(t)>=-H|P(t)>
\end{align}
with the generator
\begin{align}
\label{hamil}
H=b_1+b_L+\sum_{i=1}^{L-1}h_{i,i+1}.
\end{align}
For an explicit representation of the generator we denote
the state of a given site $i$ by the three basis vectors
\begin{align}
\label{basis}
|A>\equiv |1>=
\left(
  \begin{array}{l}
    1 \\
    0 \\
    0
  \end{array}
\right), \quad
|\emptyset>\equiv |0>=
\left(
  \begin{array}{l}
    0 \\
    1 \\
    0
  \end{array}
\right), \quad
|B>\equiv |-1>=
\left(
  \begin{array}{l}
    0 \\
    0 \\
    1
  \end{array}
\right)
\end{align}
corresponding to having an A, no particle or B at site $i$, respectively.
Let $E_k^{x,y}$ be the $3 \times 3$ matrix
with one element located at column $x$ and row $y$ equal to one. All other elements
are zero: $(E_k^{x,y})_{a,b}=\delta_{x,a}\delta_{y,b}$. The operator for
annihilation (creation) of an A particle at site $k$ is
$a_k^-=E_k^{1,2}$ ($a_k^+=E_k^{2,1}$) and for annihilation (creation) of B is
$b_k^-=E_k^{3,2}$ ($b_k^+=E_k^{2,3}$). Finally, the number operators
are $a_k=E_k^{1,1}$, $b_k=E_k^{3,3}$, $v_k=1-a_k-b_k$. This allows to compose
the generator of the process. In this representation the boundary matrices are:
\begin{align}
\label{boundary}
b_1=\alpha_A(v_1-a_1^+)+\alpha_B(v_1-b_1^+)+\gamma_A(a_1-a_1^-)+\gamma_B(b_1-b_1^-)\\
b_L=\delta_A(v_L-a_L^+)+\delta_B(v_L-b_L^+)+\beta_A(a_L-a_L^-)+\beta_B(b_L-b_L^-)
\end{align}
Hopping in the bulk between site $i$ and $i+1$ occurs according
to
\begin{multline}
\label{hopping}
h_{i,i+1}=D_A(a_iv_{i+1}+v_ia_{i+1} -a_i^-a_{i+1}^+ -a_i^+a_{i+1}^-) \\
      +D_B(b_iv_{i+1}+v_ib_{i+1} -b_i^+b_{i+1}^- -b_i^-b_{i+1}^+).
\end{multline}
The model
is now well defined. Let us proceed by deriving some equilibrium properties
of the process.

\section{Equilibrium properties}

The open system allows for particle exchange at the boundaries. The system is
ergodic and will relax to a unique stationary state determined
by the boundary rates. The stationary state $|P^*>$ does not evolve in time and must obey
\begin{align}
\label{stationary}
H|P^*>=0.
\end{align}
Let us seek a product ansatz for the equilibrium state of the form
\begin{align}
\label{equansatz}
|P*>=
\left(
  \begin{array}{l}
    a \\
    1 \\
    b
  \end{array}
\right)^{\otimes L} \frac{1}{(1+a+b)^L}.
\end{align}
The normalization factor of \eqref{equansatz} ensures conservation of probability, i.e.
ensures that the probability of finding the system in any state is one. 
Plugging the ansatz \eqref{equansatz} into \eqref{stationary} provides
a set of equations for the boundary rates and one finds $a=\frac{\alpha_A}{\gamma_A}=\frac{\delta_A}{\beta_A}$,
$b=\frac{\alpha_B}{\gamma_B}=\frac{\delta_B}{\beta_B}$. Taking into account the normalization
determines the A and B particle equilibrium densities
\begin{align}
  \label{densityA}
  \rho_A=
  \frac{\frac{\alpha_A}{\gamma_A}}{1+\frac{\alpha_A}{\gamma_A}+\frac{\alpha_B}{\gamma_B}}=
  \frac{\frac{\delta_A}{\beta_A}}{1+\frac{\delta_A}{\beta_A}+\frac{\delta_B}{\beta_B}}
\end{align}
\begin{align}
  \label{densityB}
  \rho_B=
  \frac{\frac{\alpha_B}{\gamma_B}}{1+\frac{\alpha_A}{\gamma_A}+\frac{\alpha_B}{\gamma_B}}=
  \frac{\frac{\delta_B}{\beta_B}}{1+\frac{\delta_A}{\beta_A}+\frac{\delta_B}{\beta_B}}.
\end{align}

\begin{figure}
\centerline{\includegraphics[width=14cm]{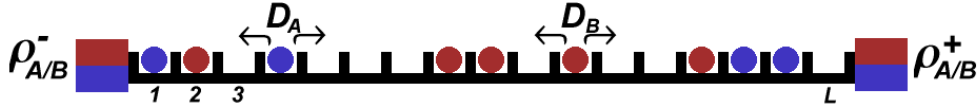}}
\caption{Three-state symmetric exclusion model with open boundaries -- reservoir picture.}
\label{latticeThreeStateReservoir}
\end{figure}

Besides giving the bulk equilibrium densities, the two equations above provide 
a recipe of how to translate the picture of inserting and deleting particles on boundary sites
into a picture of constant reservoirs at the ends, Fig.~\ref{latticeThreeStateReservoir}. In equilibrium the bulk contains
no correlations between different lattice sites and the same holds for the boundary and their
adjacent sites. Therefore, jumping from a boundary site into the chain occurs
proportional to the respective hopping rate and proportional to the single species boundary
density. Therefore,
given a set of constant boundary rates, Eqs.
\eqref{densityA} and \eqref{densityB} define the densities of a virtual particle reservoir 
at the respective boundaries. 

Note that for this interpretation the left reservoir densities $\rho_A^-$, $\rho_B^-$
do not need to be equal to their fellows on the right edge ($\rho_A^+$, $\rho_B^+$). In this
case the system evolves towards a correlated non-equilibrium stationary state with non-vanishing 
particle currents. The second and last terms of \eqref{densityA} and \eqref{densityB} 
are then stationary densities on the left and right edge of the system.  This parameterisation satisfies \eqref{densityA} and \eqref{densityB} if
\begin{align}
\label{boundaryratesleft}
\alpha_{A/B}=D_{A/B}\rho_{A/B}^-, \quad
\gamma_{A/B}=D_{A/B}(1-\rho_A^--\rho_B^-)
\end{align}
\begin{align}
\label{boundaryrateslright}
\delta_{A/B}=D_{A/B}\rho_{A/B}^+, \quad
\beta_{A/B}=D_{A/B}(1-\rho_A^+-\rho_B^+).
\end{align}

\section{Hydrodynamic limit}

The average densities $\lav a_i \rav$ and $\lav b_i \rav$ satisfy the equations of motion
$\frac{d}{dt}\lav a_i \rav=-\lav a_i H \rav$, $\frac{d}{dt}\lav b_i \rav=-\lav b_i H \rav$ (cf. \cite{Schu2001}). 
This provides the Master equations for a single site,
\begin{align}
\label{motion}
\frac{d}{dt}\lav a_i \rav
&= D_A\left( \lav a_{i-1}v_i \rav + \lav a_{i+1}v_i \rav - \lav a_i v_{i+1} \rav - \lav a_i v_{i-1} \rav \right)\\
\frac{d}{dt}\lav b_i \rav
&= D_B\left( \lav b_{i-1}v_i \rav + \lav b_{i+1}v_i \rav - \lav b_i v_{i+1} \rav - \lav b_i v_{i-1} \rav \right).
\end{align}
From now on we discuss the case of equal hopping rates $D_A=D_B=D$. Imagining the
individual particle species $A$ and $B$ to be distinguishable by a ``colour'' (in
an abstract sense) we shall refer to the total particle density (averaged over
$A$ and $B$ particles) as colourblind density.

Let us first
assume an infinite system and do not care about boundary sites.
But still, in this form the equations of motion are not integrable. Replacing
the joint probabilities by products of expectation values, according to a 
mean field ansatz which has been proven to be useful in other systems, fails.
However, an exact equation containing no correlators can be achieved from a sum of both
\begin{align}
\label{heat}
\frac{d}{dt}\left( \lav a_i \rav + \lav b_i \rav \right)
=D\left( \lav a_{i-1} \rav + \lav a_{i+1} \rav - 2\lav a_i \rav + \lav b_{i-1} \rav + \lav b_{i+1} \rav - 2\lav b_i \rav \right).
\end{align}
The right-hand side contains a second-order difference for both species individually. 
\eqref{heat} is the discrete analogue of the diffusion equation for the 
colourblind macroscopic 
profile. Introducing a lattice constant $a$ and replacing $i$ by the continuous variable $x=\frac{i}{a}$, transforms \eqref{heat} for the
hydrodynamic limit of vanishing lattice constant $a$ into
\begin{align}
\label{heatlimit}
\partial_t \left( \rho_A(x,t) + \rho_B(x,t) \right)
=D\partial_x^2(\rho_A(x,t)+\rho_B(x,t)).
\end{align}
for the macroscopic particle densities $\rho_A(x,t)$, $\rho_B(x,t)$.

Following the argument of \cite{Quastel} we make an ansatz for the dynamics of
a single particle
localized at position $x$. For a short-time region this particle acts as a tracer particle in
the background of other particles with the self-diffusion coefficient
$D_S(x,t)$. Going beyond \cite{Quastel} we argue that for the finite-size problem
with open boundaries $D_S$ is given by expression 
\begin{align}
\label{dsasb}
D_S=D\frac{1-\rho_A-\rho_B}{\rho_A+\rho_B} \frac{1}{L}.
\end{align}
derived originally for a periodic lattice \cite{vanB83}.
Additionally, the test particle is subjected to a
drift $b$ caused by the evolution of the entire system towards its stationary
state. For a good intermixed 
background one would expect the drift velocity $b$ to be the
same for both species of particles. We thus arrive at the ansatz
\begin{align}
\partial_t\rho_A(x,t)&=\partial_x^2 D_S\rho_A(x,t)-\partial_x b\rho_A(x,t) \label{ansatz1}\\
\partial_t\rho_B(x,t)&=\partial_x^2 D_S\rho_B(x,t)-\partial_x b\rho_B(x,t). \label{ansatz2}
\end{align}
The self-diffusion coefficient $D_S$ as well as the drift $b$ are functions of
$\rho_A$ and $\rho_B$ and hence, depend implicitly on $x$ and $t$. 
The drift term can be determined by using the colourblind exact result 
\eqref{heatlimit} and one finds
\begin{align}
\label{theb}
b=\frac{1}{\rho}\partial_x\left[\rho(D_S-D) \right].
\end{align}
This completes the hydrodynamic description of the two-component symmetric exclusion
process with open boundaries that we propose. A derivation of $D_S$ on a finite lattice with two different particle species \cite{Brz06} will be presented in a forthcoming paper \cite{Brza07}.

\section{Stationary state}

It is a significant property of particle systems with open boundaries 
that they can relax to a steady state with non-vanishing particle
currents. The stationary state of the colour-blind profile $\rho=\rho_A+\rho_B$ is linear with
the slope being determined by the sum of the boundary densities on the
left ($\rho^-=\rho_A^-+\rho_B^-$) and on the right edge ($\rho^+=\rho_A^++\rho_B^+$) of the system 
which is manifest by \eqref{heatlimit}, 
\begin{align}
\label{rho}
\rho=\rho^- +\left(\rho^+ - \rho^-\right)\frac{x}{L}.
\end{align}
Integrating \eqref{ansatz1} once for vanishing time derivative 
yields 
\begin{align}
\label{currents}
\frac{d}{dx} \rho_A + \left( \frac{D}{D_S}-1 \right) \frac{\rho'}{\rho}\rho_A + \frac{j_A}{D_S}=0.
\end{align}
where $j_A$ is the constant $A$-particle current. Absorbing $\rho_A^-$
into the integration constant yields the solution
\begin{small}
\begin{align}
\label{statsolution}
\rho_{A}(x)=\frac{\rho(x)}{\rho^-}\left[ -\frac{L j_A \rho^-}{D(\rho^+ -\rho^-)}
                                     +\left( \rho_A^- + \frac{L j_A \rho^-}{D(\rho^+ -\rho^-)} \right)
         \left( \frac{1-\rho(x)}{1-\rho^-} \right)^L \right] \quad \rho^+ \neq \rho^-.
\end{align}
\end{small}
The first term is linear in $x$ and describes the bulk region. The nonlinear second
term describes a boundary layer, first observed numerically in the Rubinstein-Duke
model \cite{Leeuwen2003} for different boundary rates. Our analysis shows that the
length of the boundary layer does not
scale with system size. This can be seen by rewriting \eqref{statsolution} for sufficiently large $L$ and assuming $\rho^+ > \rho^-$:
\begin{small}
\begin{align}
\label{statsolutionexp}
\rho_{A}(x)=\left(1+ \frac{\rho^+-\rho^-}{\rho^-}\frac{x}{L} \right)
            \left(-\frac{L j_A \rho^-}{D(\rho^+ -\rho^-)}
                                     +\left( \rho_A^- + \frac{L j_A \rho^-}{D(\rho^+ -\rho^-)} \right)
         e^{-\frac{\rho^+-\rho^-}{1-\rho^-}x} \right) \quad \rho^+ > \rho^-
\end{align}
\end{small}
The localization length
\begin{align}
\xi=\frac{1-\rho^-}{\rho^+-\rho^-}
\end{align}
does not depend on
the system size. In the limit of infinite $L$ 
the relative size of the boundary layer vanishes and the linear solution connects to
the reservoir densities by a jump discontinuity
at one of the edges.
For $\rho^+ > \rho^-$ the exponential in \eqref{statsolutionexp} dominates
for small $x$ and the discontinuity is located on left boundary. 
The case $\rho^+ < \rho^-$ is similar, but the discontinuity is at the right edge.

The case of equal reservoir densities $\rho_+ = \rho_-=\rho$ has to be treated separately. The self-diffusion coefficient
is constant, hence, the $b$ in \eqref{ansatz1} vanishes. Integrating 
\eqref{ansatz1} with $\rho_A(0)=\rho_A^-$ yields the
linear density profile
\begin{align}
\label{statsolution2}
\rho_{A}(x)=\rho_A^- +\left(\rho_A^+ - \rho_A^-\right)\frac{x}{L} 
\end{align}
and a similar expression for the density of $B$-particles.

\begin{figure}
\centerline{\includegraphics[width=9cm,angle=270]{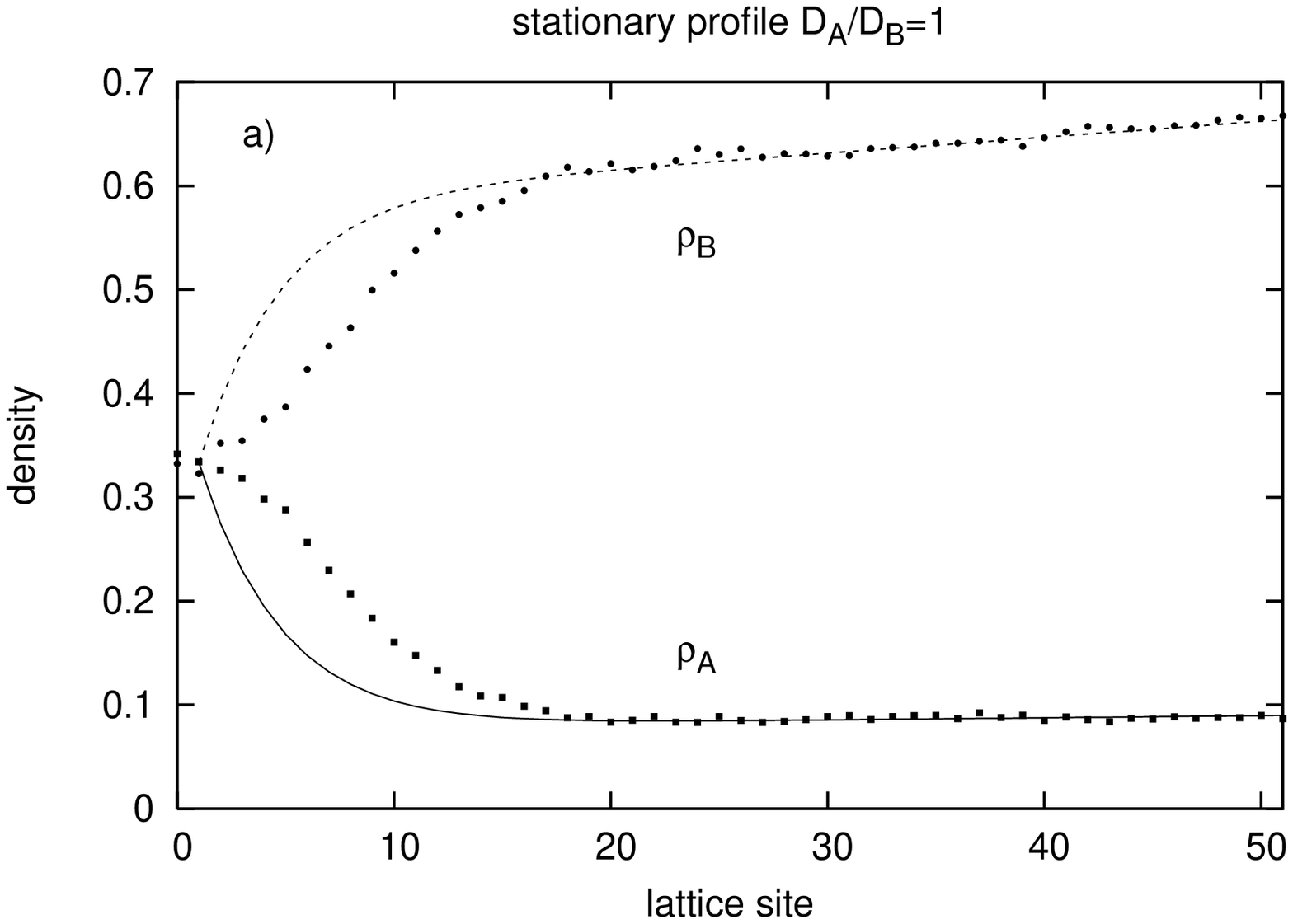}}
\centerline{\includegraphics[width=9cm,angle=270]{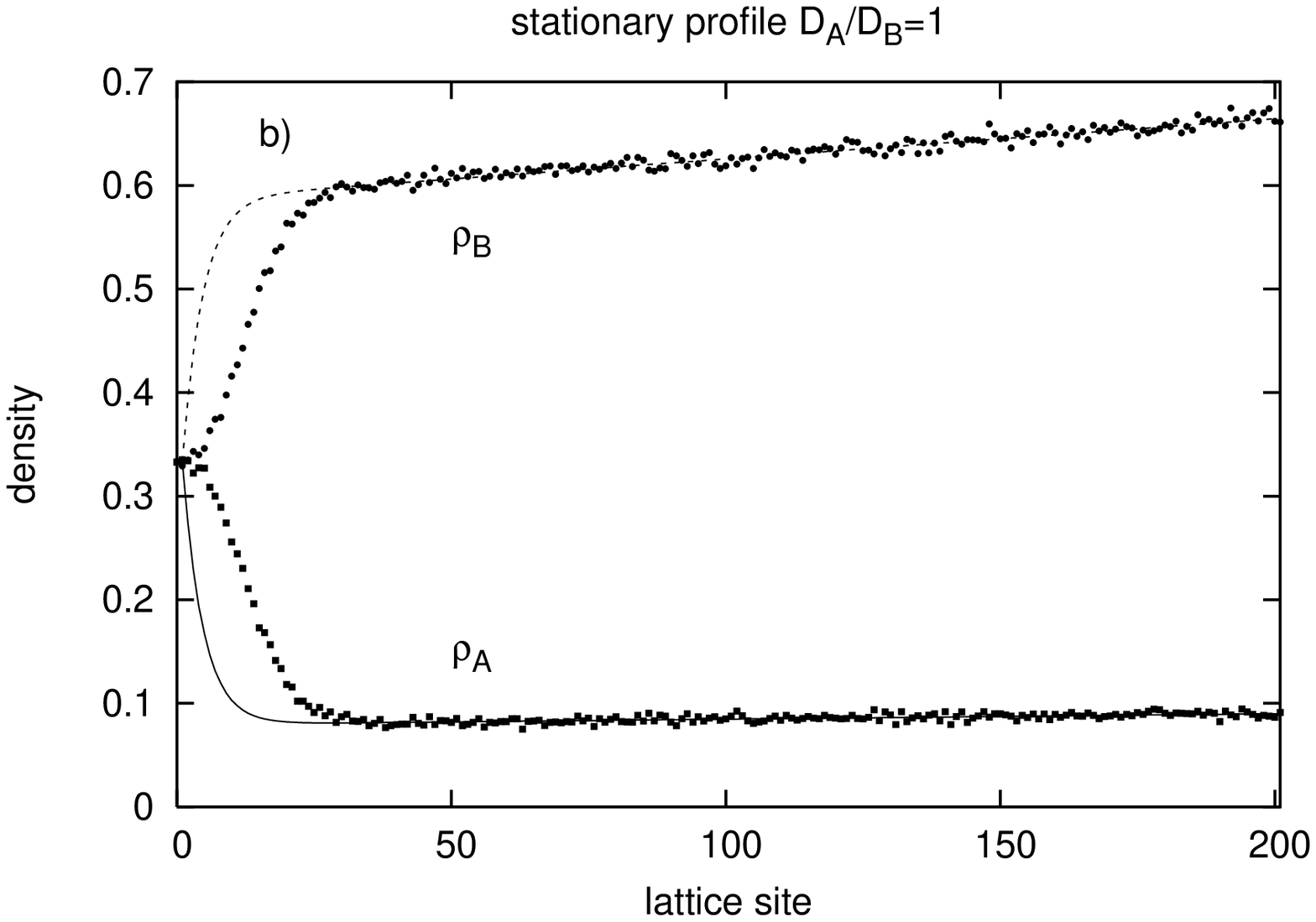}}
\caption{Stationary state with finite slope of the colour-blind density. a) $L=50$, b) $L=100$.
The boundary densities are $\rho_A^-=\rho_B^-=1/3$, $\rho_A^+=\frac{e}{1+e^{-1}+e}$ 
and $\rho_A^+=\frac{e^{-1}}{1+e^{-1}+e}$.}
\label{stat}
\end{figure}

Fig.~\ref{stat} shows the A and B particle densities obtained from Monte Carlo 
simulations (symbols) and the theoretical
stationary state solutions (solid lines). The explicit expressions for the particle
currents are given below. We apply the same set of reservoir densities $\rho_A^-=\rho_B^-=1/3$, $\rho_B^+=\frac{e}{1+e^{-1}+e}$ 
and $\rho_A^+=\frac{e^{-1}}{1+e^{-1}+e}$ used in Fig.~9 of \cite{Leeuwen2003}.
This choice is motivated by boundary rates used in the Rubinstein-Duke model 
for describing the tensile force acting at the chain ends of reptating
polymers. The different lattice sizes a) $L=50$ and b) $L=200$ demonstrate the 
finite-size character of the boundary layer. The theoretical solution does not 
contain an inflection 
point and deviates slightly from simulations in the immediate vicinity of the boundary. 
Nevertheless, an interesting observation captured by the theoretical description is confirmed. The 
solution has a minimum in the density profile (although not very pronounced in the sample A-profile of 
Fig.~\ref{stat}) and, hence, as in the Rubinstein-Duke model there exists a region where one of 
the particle currents does not follow the direction of the density gradient.

We conclude by analyzing the behaviour of the current and the mean particle density
in the system.
Using \eqref{statsolution}, \eqref{statsolution2} and taking into account the A 
particle reservoir density on the right edge gives the current
\begin{align}
\label{currentsacc}
j_A=
\begin{cases}
-\frac{D(\rho^+ - \rho^-)}{L} \frac{\frac{\rho_A^+}{\rho^+}-\frac{\rho_A^-}{\rho^-}\left(\frac{1-\rho^+}{1-\rho^-}\right)^L}
                                   {1 - \left(\frac{1-\rho^+}{1-\rho^-}\right)^L} &\rho^+ \ne \rho^-\\
-\frac{D}{L^2 \rho}(\rho_A^+ -\rho_A^-)(1-\rho)  &\rho^+ = \rho^-
\end{cases}
\end{align}
This has an interesting consequence. Considering large $L$ \eqref{currentsacc} simplifies asymptotically to
\begin{align}
\label{currentssimple}
j_A=
\begin{cases}
-\frac{D}{L} \frac{(\rho^+ - \rho^-)\rho_A^+}{\rho^+} &\rho^+>\rho^-\\
-\frac{D}{L} \frac{(\rho^+ - \rho^-)\rho_A^-}{\rho^-} &\rho^+<\rho^-\\
-\frac{D}{L^2 \rho}(\rho_A^+ -\rho_A^-)(1-\rho)  & \rho^+ = \rho^-.
\end{cases}
\end{align}
Hence, provided a finite slope of the colour-blind profile, the individual particle currents are proportional to $1/L$, as for the single-component case, whereas for 
$\rho^+=\rho^-$ the currents vanish proportional to $1/L^2$.\\

We make an interesting observation if the relation $\rho_A^- \rho_B^+ = \rho_A^+ \rho_B^-$ is satisfied.
For this particular case the individual density profiles are linear and the particle currents 
are just proportional to the respective density gradients (Fick's law). If the relation does 
not apply we observe a boundary layer inside which the current flows {\it against}
the local gradient. Here Fick's law is violated.

Finally, the mean $A$-density in the channel can be obtained
by integrating \eqref{statsolutionexp}. Asymptotically for large $L$ one finds 
from \eqref{currentssimple}
\begin{align}
\label{totaldens}
\overline{\rho_A}=
\begin{cases}
           \frac{\rho_A^+(\rho^-+\rho^+)}{2\rho^+} &\rho^+>\rho^-\\
           \frac{\rho_A^-(\rho^-+\rho^+)}{2\rho^-} &\rho^+<\rho^-\\
           \frac{\rho_A^-+\rho_A^+}{2} &\rho^+=\rho^-.
\end{cases}
\end{align}
The mean $A$-density evaluated as a function of the boundary densities may have a discontinuity. Assume
$\rho^-$ and $\rho_A^- \ne \rho_A^+$ be fixed. When taking the limit $\rho^+ \to \rho^-$ coming from 
small $\rho^+$ the total density approaches $\overline{\rho_A} \to \rho_A^-$. Taking the limit 
from the other site, $\overline{\rho_A} \to \rho_A^+$. There is a jump of the mean $A$-density when the colour-blind boundary densities become equal. Since the
colour-blind  density is the sum of $A$ and $B$ densities, this implies a jump
discontinuity also in the mean $B$-density.
Therefore there is a first-order nonequilibrium phase transition in this 
boundary-driven lattice gas model for
two-component single-file diffusion. Such a transition is not known for
boundary-driven one-component systems.

Acknowledgement: Financial support by the Deutsche Forschungsgemeinschaft is 
gratefully acknowledged. We also thank Rosemary Harris, Jörg Kärger 
and Henk van Beijeren for useful discussions.

\end{document}